\documentclass[conference]{IEEEtran}
\IEEEoverridecommandlockouts
\usepackage{cite}
\usepackage{amsmath,amssymb,amsfonts}
\usepackage{algorithm}
\usepackage{algorithmic}
\usepackage{graphicx}
\usepackage{textcomp}
\usepackage{xcolor}
\usepackage[nolist]{acronym}

\begin{acronym}
\acro{gPC}{generalised polynomial chaos}
\acro{CC}{chance constraint}
\acro{DER}{Distributed energy resource}
\acro{DR}{distributionally robust}
\acro{GS}{Gram-Schmidt}
\acro{HP}{heat pump}
\acro{EV}{electric vehicle}
\acro{PV}{photovoltaic}
\acro{DER}{distributed energy resource}
\acro{AS}{ancillary service}
\acro{TSO}{transmission system operator}
\acro{PCE}{polynomial chaos expansion}
\acro{DSO}{distribution system operator}
\end{acronym}

\usepackage{bm}

\def\BibTeX{{\rm B\kern-.05em{\sc i\kern-.025em b}\kern-.08em
    T\kern-.1667em\lower.7ex\hbox{E}\kern-.125emX}}
\begin{document}

\title{Polynomial Chaos with Dependent Random Variables for Ancillary Services Procurement \\

}

\author{\IEEEauthorblockN{Nicola Ramseyer}
\IEEEauthorblockA{\textit{DESL} \\
\textit{EPFL}\\
Lausanne, Switzerland \\
nicola.ramseyer@epfl.ch}
\and
\IEEEauthorblockN{ Matthieu Jacobs}
\IEEEauthorblockA{\textit{DESL} \\
\textit{EPFL}\\
Lausanne, Switzerland \\
matthieu.jacobs@epfl.ch}
\and 
\IEEEauthorblockN{Mario Paolone}
\IEEEauthorblockA{\textit{DESL} \\
\textit{EPFL}\\
Lausanne, Switzerland \\
mario.paolone@epfl.ch}

}

\maketitle

\begin{abstract}
This paper presents an approach for the modelling of dependent random variables using generalised polynomial chaos. This allows to write chance-constrained optimization problems with respect to a joint distribution modelling dependencies between different stochastic inputs. Arbitrary dependencies are modelled by using Gaussian Copulas to construct the joint distribution. The paper exploits the problem structure and develops suitable transformations to ensure tractability. The proposed method is applied to a probabilistic power reserve procurement problem. The effectiveness of the method to capture dependencies is shown by comparing the approach with a standard approach considering independent random variables.
\end{abstract}


\section{Introduction}
\label{sec:intro}
\Acp{DER} are increasingly participating in \acp{AS} markets. Many of these resources, such as \acp{HP}, \acp{EV} and \ac{PV} installations have uncertain capabilities. To facilitate the participation of these \acp{DER}, some \acp{TSO} already allow bids for which the full power availability is only guaranteed with a certain probability \cite{P90}. This raises new questions with respect to the amount of \acp{AS} that should be procured, as the full availability of the contracted \acp{AS} becomes uncertain. Existing works considering probabilistic reserve procurement either consider distributionally robust methods to account for uncertainties (e.g. \cite{alsox} or \cite{reserve_kuhn}) or  formulate chance constraints for the availability of different reserve bids, but do not account for the dependencies between the stochastic bid availabilities (e.g. \cite{prob_reserve}). 

Recently, \ac{gPC} has been used in power systems applications to model the propagation of uncertain input variables. Advantages include the ability to model general probability distributions and represent stochastic variables through orthogonal polynomial expansions. This approach was first used for stochastic power flow computations \cite{gpc_tillman}, with inequality constraints formulated only with respect to the expected value. Later work, such as \cite{gpc_tillman2}, introduces the use of gPC for the solution of optimal power flow problems. Typically, approaches modelling uncertainty with \ac{gPC} formulate \acp{CC} through a moment-based reformulation. For Gaussian random variables or \ac{DR} uncertainty, exact reformulations are found \cite{gpc_gaussian}. For other distributions, iterative tuning approaches for the bounds in the moment-based reformulation exist \cite{gpc_tuning} attempting to recover the desired constraint satisfaction rate. A major limitation of these approaches is, that they consider the different random input variables to be independent. This can lead to underestimating the probability of violating certain thresholds. To allow modelling dependencies between random variables, \cite{navarro2014polynomialchaosexpansiongeneral} proposes to construct an orthonormal polynomial basis using a \ac{GS} procedure. A challenge with this approach is that it leads to numerical issues due to the poor conditioning of the \ac{GS} procedure\cite{gpc_jake}. Other methods to model dependent random variables with \ac{gPC} exist, as reported for example in \cite{gpc_jake}. The Rosenblatt transformation is a mapping that transforms dependent random variables to a space of independent random variables, in which an orthonormal polynomial basis can be constructed using standard approaches. This has for example been applied to solve stochastic power flows \cite{rosenblatt_pf}. However, this approach requires to determine the inverse Rosenblatt transform for each sample used to obtain \ac{PCE} coefficients. For linear dependencies, this can be simplified to the Nataf transform \cite{nataf}, but the authors of \cite{gpc_jake} show that it performs worse than their proposed \ac{GS} procedure. Other approaches, so-called domination methods, approximate functions of dependent random variables with functions of independent random variables, but only show good performance when an accurate approximation of the orthogonal basis in terms of inner products with respect to the original joint distribution can be obtained \cite{gpc_jake}. 

None of the works above handling reserve procurement allow to consider uncertain bids exhibiting dependencies. Additionally, the proposed approaches to extend \ac{gPC} to dependent variables lead to computational issues due to error propagation and computation time.
Therefore, this paper makes the following contributions:
\begin{enumerate}
    \item An approach to model arbitrary dependencies between random variables using (Gaussian) Copulas.
    \item A tractable computation scheme for the \ac{gPC} basis, using quadrature integration and exploiting the Copula structure.
    \item A convex reformulation for probabilistic \ac{AS} procurement considering dependent random variables.
\end{enumerate}


\section{Methodology}
\label{sec:meth}

\subsection{gPC with Dependent Random Variables}
\label{sec:gpc}

Consider the $d$-dimensional random vector $\boldsymbol{\xi}=[\xi_1,\ldots,\xi_d]^\top$ with joint probability density function (PDF) $p(\boldsymbol{\xi})$ and support $\Xi$. 
As shown in previous works (e.g. \cite{navarro2014polynomialchaosexpansiongeneral}), a square-integrable random function $q(\boldsymbol{\xi})$ can be represented by a \ac{PCE}.
\begin{equation}
    q(\boldsymbol{\xi}) \cong \sum_{l=0}^{L} a_l \psi_l(\boldsymbol{\xi}).
    \label{eg:pce_sum}
\end{equation}

with the equality becoming exact as the arbitrary number of terms $L$ goes to infinity. The basis polynomials $\{\psi_l\}$ are orthonormal with respect to the inner product
\begin{equation}
\left\langle \psi_m,\psi_n \right\rangle
= \int_{\Xi} \psi_m(\boldsymbol{\xi})\,\psi_n(\boldsymbol{\xi})\,p(\boldsymbol{\xi})\,\mathrm{d}\boldsymbol{\xi}
\end{equation}

and the expansion coefficients $\{a_l\}$ are obtained by projection onto the polynomial basis.
Since this inner product corresponds to an expectation, the coefficients read
\begin{equation}
a_l
= \left\langle q,\psi_l \right\rangle
= \mathbb{E}\!\left[q(\bm{\xi})\,\psi_l(\bm{\xi})\right]
\label{eq:coeffs}
\end{equation}

The moments of the output distribution of $q(\bm{\xi})$ can be recovered from the expansion coefficients. For instance, the mean $\mu$ and variance $\sigma$ are given by
\begin{equation}
\begin{aligned}
    \mu &= a_0\\
    \sigma &\cong \left\|[a_1,\ldots,a_L]^\top \right\|_2
    \label{pce_prop}
\end{aligned}
\end{equation}

The main challenge lies in the construction of orthonormal polynomial bases with respect to an arbitrary joint PDF $p(\bm{\xi})$. For independent random variables, polynomial families associated with the Askey scheme can be employed, since they are orthogonal with respect to their corresponding marginal distributions (e.g. Hermite polynomials for Gaussian random variables).

For general dependent random variables, the basis can be constructed from a suitably chosen set of multivariate monomials of the random variables (e.g. all monomials with degree $\leq \nu$). As noted in the introduction, a GS procedure can be used for the construction, but is numerically ill-conditioned~\cite{gpc_jake}. To address this, \cite{Rahman2018_PCEdependent} provides an alternative algorithm based on a Whitening transformation, relying only in part on GS procedures.

\subsection{Joint Probability using Copulas}
\label{sec:Copula}

For practical modelling of statistically dependent random variables, the joint PDF $p(\bm{\xi})$ must be approximated in a tractable manner. This requires specifying both the marginal distributions of the individual components $f_i(\xi_i)$ as well as their dependence structure.

Copulas provide a general and flexible framework for separating the modelling of marginal distributions from the modelling of statistical dependence. Specifically, a Copula $C$ links the marginal cumulative distribution functions (CDFs) $F_i(\xi_i)$ to the joint CDF \(F(\boldsymbol{\xi})\). This link is implicitly expressed as
\begin{equation}
    F(\bm{\xi})
    = C\!\left(F_1(\xi_1),\ldots,F_d(\xi_d)\right).
    \label{eq:Copula}
\end{equation}

Sklar's Theorem states that for every multivariate CDF there exists a Copula capturing the general dependence structure given by (\ref{eq:Copula}). In practice, one of the several parametric Copula families is used. 

For the application considered in this paper, a Gaussian Copula is selected. The Gaussian Copula captures dependence through a correlation matrix $\bm{\Sigma}$ containing the pairwise correlations between the random variables. Identifying $u_i = F_i(\xi_i)$, it is defined as: 
\begin{equation}
    C_{\bm{\Sigma}}(\bm{u}) = \Phi_{\bm{\Sigma}}(\Phi^{-1}(u_1),..., \Phi^{-1}(u_d))
\end{equation}

where $\Phi_{\bm{\Sigma}}(\bm{z})$ is the multivariate normal CDF with correlation matrix $\bm{\Sigma}$ and $\Phi^{-1}(u_i)$ the inverse normal CDF. This closed-form expression allows efficient scaling to higher dimensions while maintaining computational tractability. In contrast, Archimedean Copulas provide limited flexibility in higher dimensions due to their single-parameter structure, whereas Vine Copulas can represent complex and asymmetric dependence structures but rapidly become intractable as dimensionality increases. Using a Copula, with its density expressed as $c(\bm{u})$, the joint PDF can be formulated as
\begin{equation}
    p(\bm{\xi}) = c(\bm{u})* \prod_{i=1}^{d}f_i(\xi_i)
    \label{eq:jointpdf}
\end{equation}

For a Gaussian Copula, the density is given by (\ref{eq:gausscopdens}), where $\phi_{\bm{\Sigma}}(\bm{x})$ is the multivariate normal PDF with correlation matrix ${\bm{\Sigma}}$ and $\phi(x_i)$ the normal PDF.
\begin{equation}
c_{\bm{\Sigma}}(\bm{u})
= \frac{\phi_{\bm{\Sigma}}\big(\Phi^{-1}(u_1),\ldots,\Phi^{-1}(u_d) \big)}
       {\prod_{i=1}^d \phi\big(\Phi^{-1}(u_i)\big)}
\label{eq:gausscopdens}
\end{equation}

\subsection{Quadrature Integration}
\label{sec:quadrature}

Both the construction of the orthonormal polynomial basis and the computation of the expansion coefficients require the evaluation of a large number of integrals of the form
\begin{equation}
    \int_\Xi g(\bm{\xi})\, p(\bm{\xi}) d\bm{\xi}
\end{equation}

with $g(\bm{\xi})$ an arbitrary function. In general, these integrals cannot be evaluated analytically and must be approximated numerically. Given the potentially high dimensionality of the integrals, Monte Carlo sampling is a natural choice. This, however, requires a large amount of samples to achieve sufficient accuracy and sampling errors may accumulate throughout the PCE process. Moreover, the use of importance sampling is not straightforward in this context, since the integrands encountered vary across polynomial orders and random functions.

Therefore, to enable fast and accurate numerical integration, quadrature integration schemes in combination with dimensionality-reduction strategies are used. First, the integration domain is mapped to a space with known quadrature rules. One such space is the unit cube $[0,1]^d$ on which tensor-grid Gauss-Legendre quadrature can be applied. Let
\begin{equation}
u_i = F_i(\xi_i) \in [0,1] \qquad i=1,\dots,d
\end{equation}

The Jacobian matrix of this transformation is diagonal with entries
$\partial u_i/\partial \xi_i = f_i(\xi_i)$ the marginal PDFs. Hence,
\begin{equation}
d\bm{\xi} = \frac{d\bm{u}}{\prod_{i=1}^d f_i(\xi_i)}
\end{equation}

Applying the change of variables and substituting $p(\bm{\xi})$ from (\ref{eq:jointpdf}) yields
\begin{equation}
    \int_\Xi g(\bm{\xi})\, p(\bm{\xi}) d\bm{\xi} = \int_{[0,1]^d} g\big(\bm{F}^{-1}(\bm{u})\big)\, c(\bm{u})\, d\bm{u}
\end{equation}

where $\bm{F}^{-1}(\bm{u})$ is short-hand notation for the component-wise inverse marginal CDF. Evidently, this transformation does not rely on the specific Copula used, in principle allowing any Copula family to be used for this quadrature integration scheme. However, the Copula density $c(\bm{u})$ is in general neither polynomial nor constant, meaning Gauss-Legendre quadrature will be inexact. For the specific case of Gaussian Copulas, consider a second transformation to a Gaussian latent space
\begin{equation}
z_i=\Phi^{-1}(u_i) \in (-\infty, \infty) \qquad i=1,\dots,d
\end{equation}

on which tensor-grid Gauss-Hermite quadrature is applicable. The Jacobian matrix is again diagonal with entries
\(\partial u_i/\partial z_i = \phi(z_i)\), meaning
\begin{equation}
d\bm{u} = \Big(\prod_{i=1}^d \phi(z_i)\Big)\,d\bm{z}
\end{equation}

Applying the change of variables and substituting the Gaussian Copula density from (\ref{eq:gausscopdens}) yields
\begin{equation}
\int_\Xi g(\bm{\xi})\, p(\bm{\xi}) d\bm{\xi} = \int_{\mathbb R^d} g\big(\bm{F}^{-1}(\bm{\Phi}(\bm{z}))\big)\;
\phi_{\bm{\Sigma}}(\bm{z})\, d\bm{z}
\label{eq:secondtrans}
\end{equation}
with $\bm{\Phi}(\bm{z})$ the component-wise normal CDF. 
The result in (\ref{eq:secondtrans}) features a correlated normal PDF. However, Gauss-Hermite quadrature expects to weighted by an uncorrelated normal PDF. To address this, let $\bm{\Sigma} = \bm{L} \bm{L^T}$ be the Cholesky factorisation of the correlation matrix. For a given set of quadrature nodes $\{\bm{z}_n\}$ and weights $\{w_n\}$ the linear transformation
\begin{equation}
\bm{y}_n = \bm{L}\bm{z}_n
\end{equation}
allows to ``color`` the correlation onto the quadrature nodes resulting in 
\begin{equation}
\int_\Xi g(\bm{\xi})\,p(\bm{\xi})\,d\bm{x}
\;\approx\;
\sum_{n=1}^{N} w_n \cdot g(\bm{F}^{-1}(\Phi(\bm{Lz}_n)))
\end{equation}
The number of nodes is given by $N = k^{d_I}$ with $k$ the number of evaluation points per dimension and $d_I$ the integral dimensionality. Due to this exponential scaling, dimensionality reduction strategies are necessary. Note that, even though the second transformation eliminates the non-polynomial Copula density $c(\bm{u})$ and improves numerical accuracy, there is still no exactness guarantee. Tensor-grid Gauss–Hermite quadrature is exact only for integrands that are polynomials up to degree \(2k-1\) in each variable weighted by a normal PDF, but $g(\bm{F}^{-1}(\bm{\Phi}(\bm{Lz})))$ does not necessarily behave like a polynomial in Gaussian latent space. For a comprehensive treatment of quadrature rules and their exactness properties, see classical texts such as \cite{Arthur1986MethodsON}.\\

\subsubsection{Numerical Improvements}

The dimensionality-reduction strategy used to make quadrature integration tractable stems from two key observations: First, as noted in \cite{Rahman2018_PCEdependent}, inner products between polynomials can be expanded to a sum of monomial expectation values. For instance, let $P_m(\bm{\xi}),P_n(\bm{\xi})$ be two arbitrary polynomials constructed from the set of multivariate monomials of degree $\leq \nu$ $\{1,\xi_1,\ldots,\xi_d^\nu\}$. Then the inner product
\begin{align}
    \left\langle P_m,P_n \right\rangle & = d_0\!\left\langle 1,1 \right\rangle + d_1\! \left\langle 1,\xi_1 \right\rangle + \ldots + d_D\!\left\langle \xi_d^\nu,\xi_d^\nu \right\rangle \notag\\
    & = e_0 + e_1 \mathbb{E}[\xi_1] + \ldots + e_E \mathbb{E}[\xi_d^{2\nu}]
    \label{eq:monoexpand}
\end{align}

where $d_i, e_i$ are the appropriate coefficients. Second, the expectation values of monomials only depend on dimensions appearing in the monomial, for instance
\begin{equation}
\mathbb{E}[\xi_1\xi_2] = \int_{\Xi_{1,2}} \xi_1\xi_2 \, p_{1,2}(\xi_1,\xi_2)\, d\bm{\xi}_{1,2}.
\end{equation}

where $\Xi_{1,2}$ is the corresponding support and $p_{1,2}(\xi_1,\xi_2)$ the joint PDF of the appearing random variables. For elliptic Copulas, like the Gaussian Copula, such lower dimensional joint PDFs have identical structure to the full joint PDF defined in (\ref{eq:jointpdf}) and are therefore easy to recover. For other Copulas, they might need to be computed separately.

Now, instead of directly applying an orthonormalization scheme, the expectation values of all pairwise products of monomials in the chosen set can be evaluated beforehand. When applying the orthonormalization scheme, the inner products can be expanded as in (\ref{eq:monoexpand}) and the expectation values substituted. Crucially, precomputing expectation values can be done in a parallel fashion and naturally limits the maximum integral dimension to be evaluated to $2\nu$, as the highest dimensional expectation values to be evaluated are of the form $\mathbb{E}[\xi_1 \xi_2 \ldots \xi_{2\nu}]$. This decouples the integral dimensionality $d_I$ from the random variable dimension $d$. In the case where multiple random functions $q_i(\bm{\xi_i})$ only depending on subsets $\bm{\xi_i}$ of the random vector are expanded to the same polynomial basis, a similar approach can be used for the expansion coefficients. Here the expectation values $\mathbb{E}[q_i(\bm{\xi_i})],\ldots,\mathbb{E}[q_i(\bm{\xi_i})\xi_d^{\nu}]$ can be computed beforehand with reduced dimension. Note, that the reduction in this case is dependent on the dimensionality of the random functions $q_i(\bm{\xi_i})$. The implementation of the quadrature integration procedure using Gaussian Copulas is summarized in Algorithm~\ref{alg:quadrature_integration}.

Finally, the set of monomials from which the polynomial basis is constructed can be reduced to limit the computational cost. For instance, assume the random functions $q_i(\bm{\xi_i})$ are models of PV power output at locations $i$ with $\bm{\xi_i}$ the local weather conditions like irradiation and temperature. Physically irrelevant monomials, such as higher-order products of the irradiations at different locations, can be dropped from the set without significantly affecting the PCE accuracy. This approach is conceptually similar to sparse PCE, which discards basis polynomials with negligible influence on the expansion output \cite{BlatmanSudret2008SparsePCE}.

\section{Use Case}
\label{sec:use case}
Reserve power, procured from AS markets, is activated whenever the grid frequency exceeds or dips below certain thresholds to ensure stable frequency levels. Given the critical nature of reserve activation, the risk of under-delivery must be kept to a minimum. This need for security can lead to high procurement costs. A recent initiative to keep economic efficiency is the Allocation of Cross-zonal Capacity and Procurement of aFRR Cooperation Agreement (ALPACA) which co-optimizes the procurement of reserves across borders of Austria, Czechia and Germany. To validate the formalism introduced in this paper, a simple cross-border joint procurement optimization with uncertain bids and chance constraints on the available power is solved.


Formally, let $X$ and $Y$ be two zones sharing a common AS market with a set of bids $E_i(\bm{\xi})$ and costs $\gamma_i$. Bids $i = {1,\ldots, N_X}$ are located in zone $X$ and $i = {N_X + 1,\ldots, N_X + N_Y}$ are located zone Y. The decision variables $x_i, y_i \in [0,1]$ determine the procured share of each bid. The objective is to minimize total cost, while both zones meet their reserve needs $R^{(X)}, R^{(Y)}$ and stay below exchange limits of the interconnecting tie lines $T_{X \to Y}, T_{Y \to X}$ with $99\%$ probability. The value of this probability threshold is inspired from Art 157. (h)-(i) in \cite{EU2017Reg1485}. Expressing the problem in standard form
\begin{equation}
\begin{aligned}
\min_{\bm{x},\bm{y}} \quad & \bm{\gamma}^\top (\bm{x} + \bm{y})\\
\text{s.t.} \quad & 0 \le \bm{x} + \bm{y} \le 1 \\
& \mathbb{P}\left( \sum_{i=1}^{N_X+N_Y} x_i E_i(\bm{\xi}) \geq R^{(X)} \right) \geq 0.99\\
& \mathbb{P}\left( \sum_{i=1}^{N_X+N_Y} y_i E_i(\bm{\xi}) \geq R^{(Y)}\right) \geq 0.99\\
& \mathbb{P}\left( \sum_{i=1}^{N_X} y_i E_i(\bm{\xi}) \leq T_{X\to Y}\right) \geq 0.99 \\
& \mathbb{P}\left( \sum_{i=N_X+1}^{N_X+N_Y} x_i E_i(\bm{\xi}) \leq T_{Y\to X}\right) \geq 0.99 \\
\end{aligned}
\label{eq:ccnonconv}
\end{equation}

with the first constraint ensuring each bid is procured at most once. To solve this problem, the chance-constraints have to be reformulated. For this, note that arbitrary weighted sums of gPC expansions result in a new gPC expansion
\begin{algorithm}[t]
\caption{Quadrature Integration with Dimension Reduction for a Gaussian Copula}
\label{alg:quadrature_integration}
\begin{algorithmic}[1]

\REQUIRE Integrand function $g(\bm{\xi})$, correlation matrix $\bm{\Sigma} \in \mathbb{R}^{d \times d}$, number of evaluation points per dimension $k$
\ENSURE Numerical integral value $I$

\STATE Determine the effective dimension $d_I$ of $g(\bm{\xi})$, 
\STATE Extract the submatrix $\bm{\Sigma}_{d_I} \in \mathbb{R}^{d_I \times d_I}$ corresponding to the relevant dimensions
\STATE Compute the Cholesky factorization $\bm{LL}^\top = \bm{\Sigma}_{d_I}$
\STATE Determine the $N=k^{d_I}$ tensor-grid Gauss-Hermite quadrature nodes $\{\bm{z_n}\}$ and weights $\{w_n\}$
\STATE $I \leftarrow 0$
\FOR{$n = 1$ to $N$}
    \STATE $I \leftarrow I + w_n \cdot g(\bm{F}^{-1}(\bm{\Phi}(\bm{Lz_n})))$
\ENDFOR
\STATE \RETURN $I$

\end{algorithmic}
\end{algorithm}

\begin{figure}
    \centering
    \includegraphics[width=0.7\linewidth]{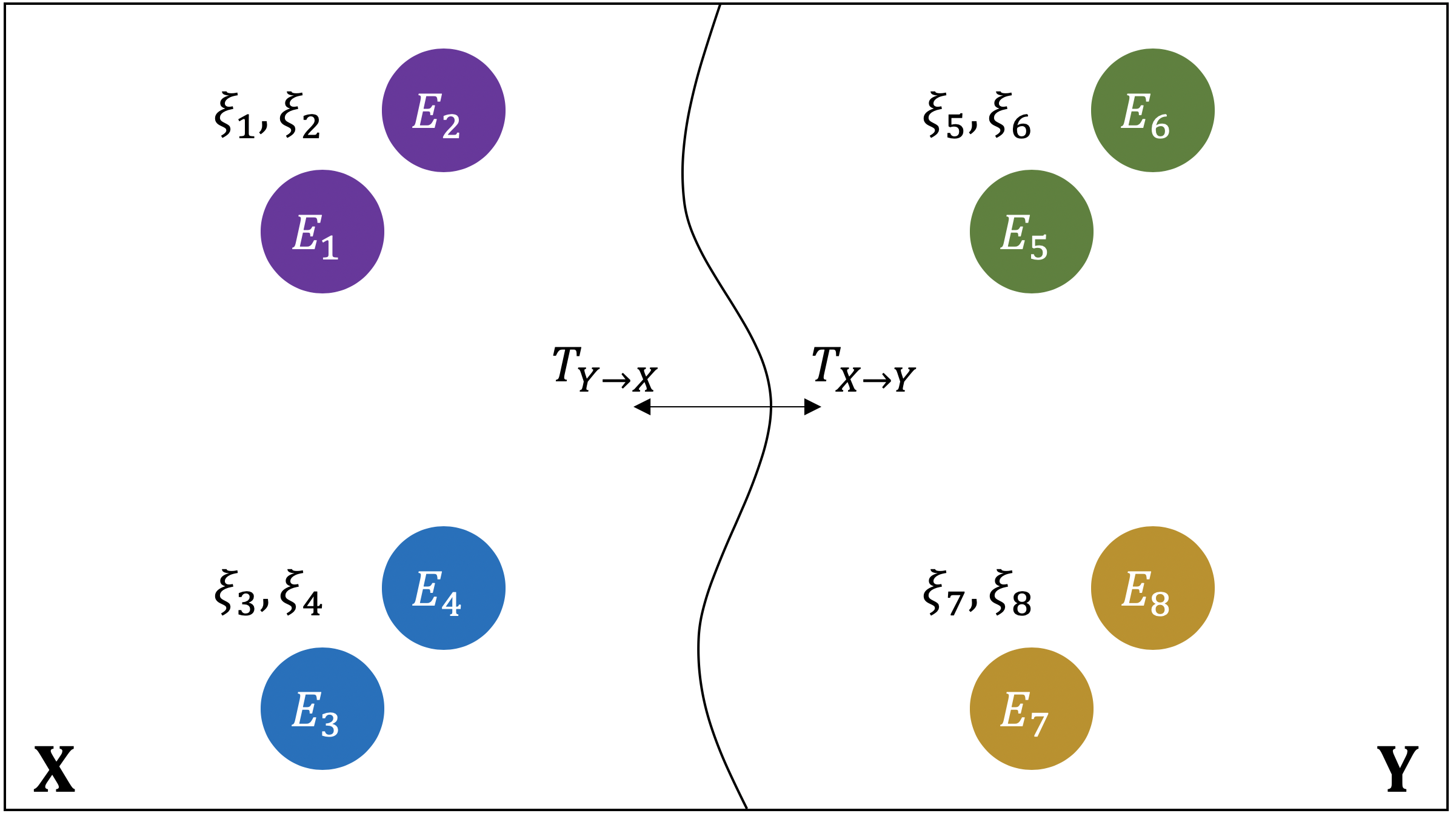}
    \caption{Illustration of the use case with four bids per zone. Two bids each are distributed across four locations with each two local random variables describing the temperature and irradiation.}
    \label{fig:visuse}
\end{figure}

\begin{equation}
\begin{aligned}
    \sum_{j=1}^{Z} z_j E_j(\bm{\xi}) &= \sum_{j=1}^{Z} z_j \sum_{l=0}^{L} a_l^j \psi_l(\bm{\xi})= \sum_{l=0}^{L} \Bar{a}_l \psi_l(\bm{\xi})
\end{aligned}
\end{equation}

where $\Bar{a}_l = \sum_{j=1}^{Z} z_j a_l^j$. Using the properties in (\ref{pce_prop}), the mean and variance of such a weighted sum can be expressed as functions of the weights 
\begin{equation}
\begin{aligned}
    \mu &= \sum_{j=1}^{Z} z_j a_0^j \\ 
    \sigma &= \left\| \begin{bmatrix} a_1^{1} & \ldots &  a_1^{Z}\\ \vdots  &\ddots   & \vdots  \\ a_L^{1} & \ldots & a_L^{Z} \end{bmatrix}
    \begin{bmatrix} z_1 \\ \vdots \\ z_{Z} \end{bmatrix}\right\|_2 \label{eq:weightmean} = \left\| (\bm{Az})_{Z}^1 \right\|_2
\end{aligned}
\end{equation}

where $(\cdot)^1_Z$ indicates the index range of the matrix columns and vector. This allows a convex reformulation of the chance-constraints using quantile approximation, for instance 
\begin{equation}
\begin{aligned}
    \mathbb{P} \left( \sum_{i=1}^{N_X+N_Y} x_i E_i(\bm{\xi}) \geq R^{(X)} \right) \geq 0.99 \Leftrightarrow\\
    \sum_{i=1}^{N_X+N_Y} x_i a_0^i + \lambda_1 \left\|(\bm{Ax})^1_{N_X+N_Y} \right\|_2 \geq R^{(X)}
\end{aligned}
\end{equation}

where $\lambda_1$ is the quantile factor for which $\mu + \lambda_1*\sigma$ corresponds to the $1^{st}$ percentile of the distribution. For normal distributions $\lambda_1,\lambda_{99} = \mp 2.326$. Applying this reformulation to all chance-constraints in problem (\ref{eq:ccnonconv}) results in the convex optimization problem (\ref{eq:ccconv}).
\begin{equation}
\begin{aligned}
\min_{\bm{x},\bm{y}} \quad & \bm{c^T} (\bm{x} + \bm{y})\\
\text{s.t.} \quad & 0 \le \bm{x} + \bm{y} \le 1 \\
& \sum_{i=1}^{N_X+N_Y} x_i a_0^i + \lambda_1 \left\|(\bm{Ax})_{N_X+N_Y}^1 \right\|_2 \geq R^{(X)}\\
& \sum_{i=1}^{N_X+N_Y} y_i a_0^i + \lambda_1 \left\|(\bm{Ay})_{N_X+N_Y}^1 \right\|_2 \geq R^{(Y)}\\
& \;\;\; \sum_{i=1}^{N_X} \;\;\; y_i a_0^i + \lambda_{99} \left\|(\bm{Ay})_{N_X}^1 \right\|_2 \leq T_{X \to Y} \\
& \sum_{i=N_X + 1}^{N_X + N_Y} x_i a_0^i + \lambda_{99} \left\|(\bm{Ax})_{N_X+N_Y}^{N_X+1} \right\|_2 \leq T_{Y \to X}  \\
\end{aligned}
\label{eq:ccconv}
\end{equation}

We consider a concrete example with $N_X, N_Y = 4$ bids per zone and $d=8$ random variables and analyze two distinct cases. The bids are distributed across four locations, with two bids at each location depending on two random variables that represent local conditions. Fig. \ref{fig:visuse} illustrates the use case.\\

\subsubsection{Normal marginals}

\begin{figure}
    \centering
    \includegraphics[width=\linewidth]{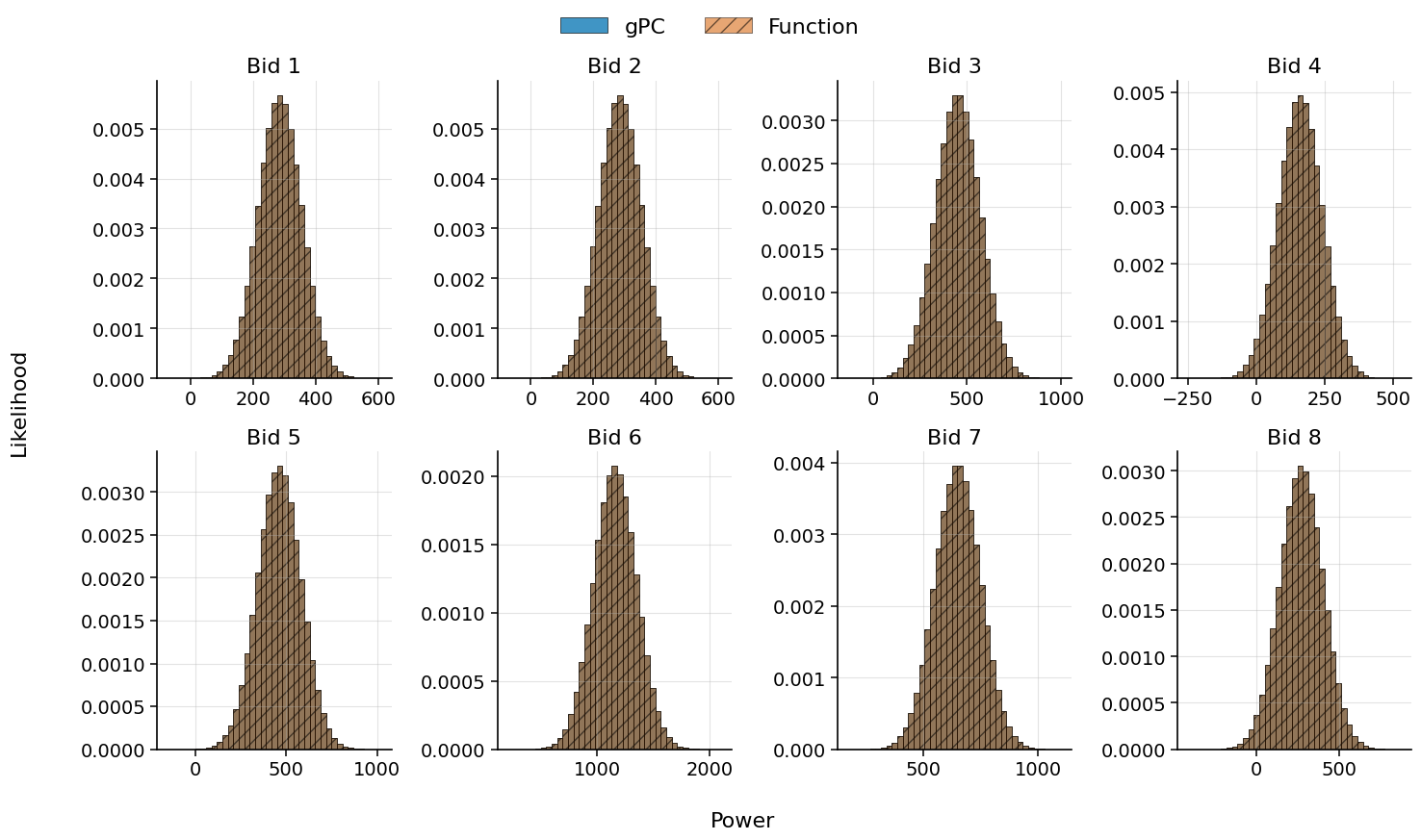}
    \caption{Sampled bid function distributions computed both with the bid functions and gPC expansions for the eight bids in the case of normally distributed marginals and linear bid functions.}
    \label{fig:normlinbid}
\end{figure}

As $\lambda$ values are known exactly for normal distributions, the full process is first validated with arbitrary normally distributed marginals and linear bid functions, together with an arbitrary correlation matrix $\bm{\Sigma}$.
The polynomial basis is constructed from 9 monomials (all monomials of degree $\leq1$). The quality of the gPC expansion is checked by drawing random samples from the Copula and evaluating both the bid functions and the expansion. The resulting distributions are displayed in Fig. \ref{fig:normlinbid}. As constructed, the bid power distributions are indeed normally distributed. Moreover, the gPC expansion accurately approximates the bid functions. The maximum numerical error of the expansion is found to be on the order of $10^{-7}\%$, making the two distributions indistinguishable.

\begin{figure}
    \centering
    \includegraphics[width=\linewidth]{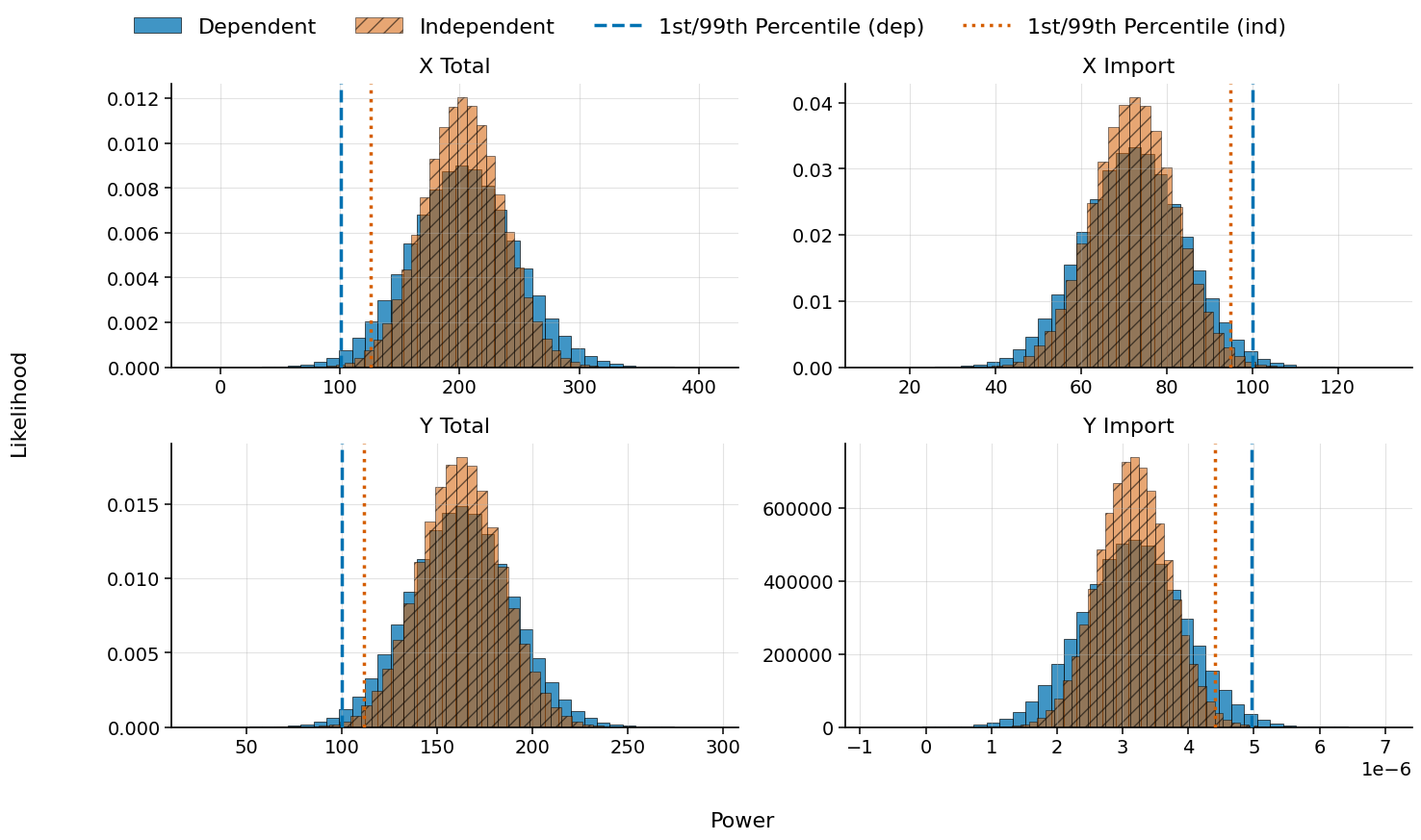}
    \caption{Sampled procured power distributions computed with the gPC expansion in the case of normally distributed marginals. For comparison the PDFs both considering and not considering dependence are sampled. The empirical $1^{st}$ and $99^{th}$ percentiles are indicated for verification of the chance-constraints.}
    \label{fig:normlinopt}
\end{figure}

Problem (\ref{eq:ccconv}) is solved with parameters $R^{(X)}$, $R^{(Y)}$, $T_{X \to Y}$, $T_{Y \to X} = 100$ and $\gamma_{1-8} = 1$ to get optimal $\bm{x},\bm{y}$. The chance-constraints in (\ref{eq:ccnonconv}) are validated by sampling both from the Copula and from the marginal distributions assuming independence. The resulting distributions are shown in Fig.~\ref{fig:normlinopt}. As expected, the optimization problem tightly satisfies the chance-constraints only for the dependent case. This result shows that the presented formalism works as intended.\\

\subsubsection{Arbitrary marginals}

In general, bid functions are not linear in the random variables and the marginals are not normally distributed. For example, solar irradiance uncertainty is frequently modelled using Beta distributions due to its bounded support. For validation in a more realistic setting, the bid functions are replaced by simple polynomials and the marginals by representative Beta distributions. 

\begin{figure}
    \centering
    \includegraphics[width=\linewidth]{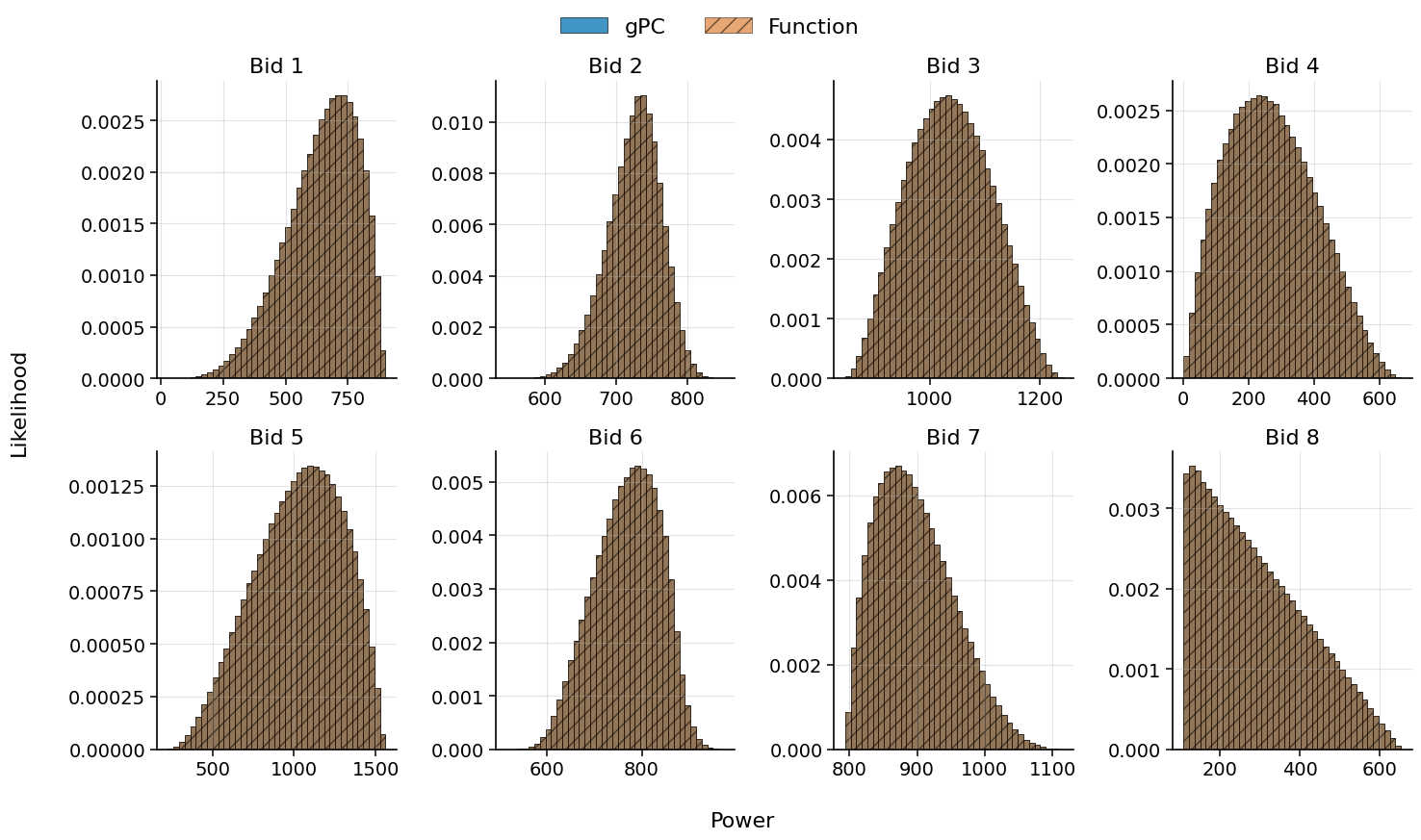}
    \caption{Sampled bid function distributions computed both with the bid functions and gPC expansions for the eight bids in the case of arbitrarily distributed marginals and polynomial bid functions.}
    \label{fig:betabid}
\end{figure}

The polynomial basis is constructed from 21 relevant monomials. The computation of the basis and all coefficients is achieved in roughly 2 minutes total with a non-parallelised implementation and $k=15$ quadrature points per dimension. Validating the gPC expansions with sampling results in the bid power distributions shown in Fig. \ref{fig:betabid}. The maximum expansion error is found to be on the order of $10^{-6}\%$, again with the two distributions indistinguishable.

\begin{figure}
    \centering
    \includegraphics[width=\linewidth]{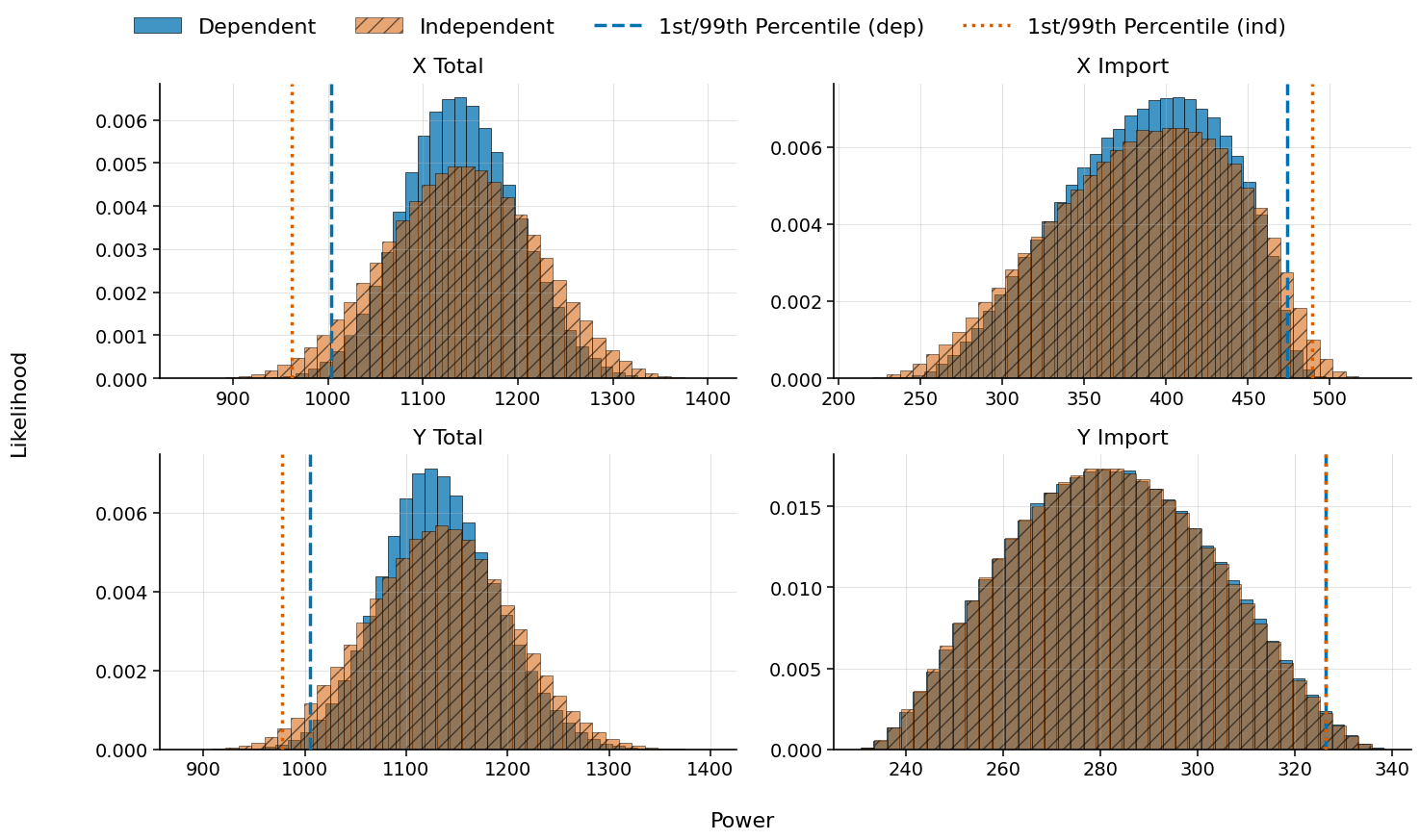}
    \caption{Sampled procured power distributions computed with the gPC expansion in the case of arbitrarily distributed marginals. For comparison the PDFs both considering and not considering dependence were sampled. The empirical $1^{st}$ and $99^{th}$ percentiles are indicated for verification of the chance-constraints.}
    \label{fig:betaopti}
\end{figure}

Problem (\ref{eq:ccconv}) is solved with parameters $R^{(X)}$, $R^{(Y)} = 1000$, $T_{X \to Y}$, $T_{Y \to X} = 500$ and $\gamma_{1-8} = 1$. The validation of constraints satisfaction through sampling results in the distributions shown in Fig.~\ref{fig:betaopti}. Due to the non-linearity of the bid functions and the non-normal, non-symmetric marginals, the constraints with $\lambda$ values for normal distributions are no longer met tightly. 
Note, that the overall computation time is dominated by the PCE process, while the choice of parameters has no noticeable impact, besides determining the feasibility of the problem. 


\section{Conclusion}
\label{sec:conclusion}

In this paper, we develop a tractable framework for modelling correlated random variables using \ac{PCE} combined with Gaussian Copulas. The approach enables to reformulate chance constraints on weighted sums of random functions as convex constraints. Several ideas are introduced to improve computational efficiency, including transformations for quadrature integration of Gaussian Copulas as well as precomputing expectation values for parallelization and effective dimensionality reduction. The framework is validated on a numerical joint procurement problem with eight random variables and bids. First, using normally distributed marginals and linear bid functions, we show that the reformulation tightly meets constraints for a correlated joint PDF. Then, for a more representative example with polynomial bid functions and Beta distributed marginals, a 21-dimensional orthonormal polynomial basis and corresponding PCE coefficients could be computed within two minutes. The optimization closely satisfies the chance constraints when using Gaussian quantile factors. In addition to the methodological contributions, the proposed framework has relevance for system operators. For instance, \acp{TSO} engaged in interzonal AS provision or \acp{DSO} participating in reserve procurement can potentially benefit from the ability to model correlated stochastic resources in a tractable optimization setting. Key limitations include the reliance of dimensionality reduction techniques on the dimensionality of the random functions as well as the need for quantile factors $\lambda$ in the exact reformulation of chance constraints. While $\lambda$ is known for normal distributions and conservative robust bounds are available, its exact value remains generally unknown for arbitrary distributions. Future work will focus on extending the framework to time-coupled problems, accounting for correlations across time steps. In addition, the formulation of joint chance constraints will be investigated.

\bibliographystyle{IEEEtran}
\bibliography{main.bib}

@misc{navarro2014polynomialchaosexpansiongeneral,
      title={Polynomial Chaos Expansion for general multivariate distributions with correlated variables}, 
      author={Maria Navarro and Jeroen Witteveen and Joke Blom},
      year={2014},
      eprint={1406.5483},
      archivePrefix={arXiv},
      primaryClass={math.NA},
      url={https://arxiv.org/abs/1406.5483}, 
}

@INPROCEEDINGS{gpc_tillman,
  author={Mühlpfordt, Tillmann and Faulwasser, Timm and Hagenmeyer, Veit},
  booktitle={2016 IEEE Conference on Control Applications (CCA)}, 
  title={Solving stochastic {AC} power flow via polynomial chaos expansion}, 
  year={2016},
  volume={},
  number={},
  pages={70-76}
}

@ARTICLE{gpc_tuning,
  author={Koirala, Arpan and Van Acker, Tom and Hashmi, Md Umar and D'hulst, Reinhilde and Van Hertem, Dirk},
  journal={IEEE Transactions on Power Systems}, 
  title={Chance-Constrained Optimization Based PV Hosting Capacity Calculation Using General Polynomial Chaos}, 
  year={2024},
  volume={39},
  number={1},
  pages={2284-2295}}

@ARTICLE{gpc_gaussian,
  author={Mühlpfordt, Tillmann and Roald, Line and Hagenmeyer, Veit and Faulwasser, Timm and Misra, Sidhant},
  journal={IEEE Transactions on Power Systems}, 
  title={Chance-Constrained AC Optimal Power Flow: A Polynomial Chaos Approach}, 
  year={2019},
  volume={34},
  number={6}}

@INPROCEEDINGS{gpc_tillman2,
  author={Mühlpfordt, Tillmann and Faulwasser, Timm and Roald, Line and Hagenmeyer, Veit},
  booktitle={2017 IEEE 56th Annual Conference on Decision and Control (CDC)}, 
  title={Solving optimal power flow with non-Gaussian uncertainties via polynomial chaos expansion}, 
  year={2017},
  volume={},
  number={},
  pages={4490-4496}}

@misc{P90,
      title={{Leveraging P90 Requirement: Flexible Resources Bidding in Nordic Ancillary Service Markets}}, 
      author={Peter A. V. Gade and others},
      year={2024},
      eprint={2404.12807},
      archivePrefix={arXiv},
      primaryClass={eess.SY},
}

@article{Rahman2018_PCEdependent,
  author    = {Sharif Rahman},
  title     = {A Polynomial Chaos Expansion in Dependent Random Variables},
  journal   = {Journal of Mathematical Analysis and Applications},
  volume    = {464},
  number    = {1},
  pages     = {749--775},
  year      = {2018},
  doi       = {10.1016/j.jmaa.2018.04.032}
}

@article{gpc_jake,
title = {Polynomial chaos expansions for dependent random variables},
journal = {Computer Methods in Applied Mechanics and Engineering},
volume = {351},
pages = {643-666},
year = {2019},
issn = {0045-7825},
doi = {https://doi.org/10.1016/j.cma.2019.03.049},
author = {John D. Jakeman and Fabian Franzelin and Akil Narayan and Michael Eldred and Dirk Plfüger}
}

@INPROCEEDINGS{rosenblatt_pf,
  author={Ye, Ketian and Zhao, Junbo and Yang, Rui and Zhang, Yingchen and Liu, Xiaodong},
  booktitle={2020 52nd North American Power Symposium (NAPS)}, 
  title={A Generalized Copula-Polynomial Chaos Expansion for Probabilistic Power Flow Considering Nonlinear Correlations of PV Injections}, 
  year={2021},
  volume={},
  number={},
  pages={1-6},
  doi={10.1109/NAPS50074.2021.9449783}}

@article{nataf,
title = {Multivariate distribution models with prescribed marginals and covariances},
journal = {Probabilistic Engineering Mechanics},
volume = {1},
number = {2},
pages = {105-112},
year = {1986},
issn = {0266-8920},
doi = {https://doi.org/10.1016/0266-8920(86)90033-0},
author = {Pei-Ling Liu and Armen {Der Kiureghian}}
}

@article{Arthur1986MethodsON,
  title={Methods of numerical Integration},
  author={D. W. Arthur and Philip J. Davis and Philip Rabinowitz},
  journal={Academic Press},
  year={1984},
}

@misc{EU2017Reg1485,
  author = {{European Commission}},
  title = {Commission Regulation (EU) 2017/1485 of 2 August 2017 establishing a guideline on electricity transmission system operation},
howpublished = {OJ L 220, 25.8.2017, p. 1--120},
}

@ARTICLE{alsox,
  author={Wen, Yilin and Guo, Yi and Hu, Zechun and Hug, Gabriela},
  journal={IEEE Transactions on Power Systems}, 
  title={Multiple Joint Chance Constraints Approximation for Uncertainty Modeling in Dispatch Problems}, 
  year={2025},
  volume={40},
  number={1},
  pages={662-675},
  doi={10.1109/TPWRS.2024.3415652}}

@article{BlatmanSudret2008SparsePCE,
  author  = {Blatman, G{\'e}raud and Sudret, Bruno},
  title   = {Sparse polynomial chaos expansions and adaptive stochastic finite elements using a regression approach},
  journal = {Comptes Rendus M{\'e}canique},
  volume  = {336},
  number  = {6},
  pages   = {518--523},
  year    = {2008},
  doi     = {10.1016/j.crme.2008.02.013}
}

@article{reserve_kuhn,
title = {Energy and reserve dispatch with distributionally robust joint chance constraints},
journal = {Operations Research Letters},
volume = {49},
number = {3},
pages = {291-299},
year = {2021},
issn = {0167-6377},
doi = {https://doi.org/10.1016/j.orl.2021.01.012},
author = {Christos Ordoudis and Viet Anh Nguyen and Daniel Kuhn and Pierre Pinson}
}

@ARTICLE{prob_reserve,
  author={Guo, Zhenwei and Pinson, Pierre and Chen, Shibo and Yang, Qinmin and Yang, Zaiyue},
  journal={IEEE Transactions on Smart Grid}, 
  title={Chance-Constrained Peer-to-Peer Joint Energy and Reserve Market Considering Renewable Generation Uncertainty}, 
  year={2021},
  volume={12},
  number={1},
  pages={798-809},
  doi={10.1109/TSG.2020.3019603}}
\end{document}